\documentclass[12pt,preprint]{aastex}
\usepackage{amsmath}
\begin{document}

\title{Local helioseismology and correlation tracking
analysis of surface structures in realistic
simulations of solar convection}

\author{Dali Georgobiani, Junwei Zhao, Alexander G. Kosovichev}
\affil{Hansen Experimental Physics Laboratory, Stanford
       University, Stanford, CA, 94305}

\author{David Benson, Robert F. Stein}
\affil{Physics and Astronomy
       Department, Michigan State University, East Lansing, MI 48824}

\author{{\AA}ke Nordlund}
\affil{Niels Bohr Institute, Copenhagen University,
       Juliane Maries Vej 30, DK--2100 K{\o}benhavn {\O}, Denmark}


\begin{abstract}
We apply time-distance helioseismology, local correlation tracking
and Fourier spatial-temporal filtering methods to realistic
supergranule scale simulations of solar convection and compare the
results with high-resolution observations from the SOHO Michelson
Doppler Imager (MDI).  Our objective is to investigate the surface
and sub-surface convective structures and test helioseismic
measurements.  The size and grid of the computational domain are 
sufficient to resolve various convective scales from granulation 
to supergranulation. 
The spatial velocity spectrum is approximately a power law for scales
larger than granules, with
a continuous decrease in velocity amplitude with increasing size.
Aside from granulation no special scales exist, although a
small enhancement in power at supergranulation scales can be seen.
We calculate the time-distance diagram for $f$- and $p$-modes
and show that it is consistent with the SOHO/MDI observations.
From the simulation data we calculate travel time maps for surface 
gravity waves ($f$-mode).  We also apply correlation tracking to 
the simulated vertical velocity in the photosphere to calculate the
corresponding horizontal flows.  We compare both of these to the
actual large-scale (filtered) simulation velocities.
All three methods reveal similar large scale convective patterns and
provide an initial test of time-distance methods.

\end{abstract}

\section{Introduction}
\label{intro} Local helioseismology uses the observed oscillation
signals to probe the near surface structure of the Sun to determine
sound speed, flow velocities and magnetic field structures. One of
several methods used in this field is called the time-distance
helioseismology. It measures the time taken by a wave packet to
travel from one point on the solar surface to another
\citep{1993Natur.362..430D, 1997scor.proc..241K}. The travel time
for a wave depends on the wave speed and the flow
velocities along the ray paths. These effects can be separated by
measuring the wave travel time for waves propagating in opposite directions
along the same ray paths. Inversion methods applied to infer the
properties of subsurface convection and structures are based on
various approximations, such as a geometric acoustic ray approximation
\citep{2000SoPh..192..159K, 2001ApJ...557..384Z}, Fresnel-zone
approximation \citep{2001ApJ...553L.193J, 2004ApJ...607..554C}
and Born-approximation \citep{2002ApJ...571..966G}.
Unfortunately, until now, there has been no direct verification of
these methods \citep{2003soho...12..319J,2004soho...14..172W}.

Realistic 3D simulations of solar convection
\citep{{2000SoPh..192...91S},{Benson2006}} provide a way to
evaluate the accuracy and consistency of time-distance and other
local helioseismology techniques. These simulations provide a
model of convective motions in the upper convection zone
and possesses a rich spectrum of $f$- and $p$-mode oscillations excited
by the turbulent convection.  They can be used for
testing helioseismology methods. Therefore, it is very important
to apply some of the existing local helioseismology methods to the
simulated solar convection, and check whether the results are
consistent with the subsurface properties present in
the simulated flows. In addition, the comparison of the
oscillation properties obtained from the numerical model, such as
the  velocity spectrum, $k-\omega$ diagram and time-distance
(cross-covariance) diagram, with observations provides validation
for the simulation model.

Previously, the realistic simulations of solar convection have been
used to study the excitation mechanism of solar and stellar
oscillations \citep{2004SoPh..220..229S}, and some properties of
oscillation modes, such as the line asymmetry
\citep{2000ApJ...530L.139G}. However, these simulations were carried
out for small, granulation-size, computational domains. Recently,
with the progress of parallel supercomputing it became possible to
substantially expand the computational domain, both horizontally and
in depth, and to simulate the multi-scale solar convection, from
granulation to supergranular scales \citep{Benson2006}. These
simulations allow us for the first time to investigate the
properties of large-scale convection by applying helioseismology and
other methods used to analyze solar observations, and thus test
these methods and also evaluate how close the realistic simulations
are to the real Sun. Such tests are particularly important for the
local helioseismology methods, which are based on simplified models
of wave propagation.

The goal of this paper is to investigate the basic helioseismic
properties of the large-scale 3D simulations (the oscillation power
spectrum and wave travel times) and compare these with the
high-resolution observations from SOHO/MDI
\citep{1995SoPh..162..129S}.  An initial time-distance 
analysis is carried out for surface gravity waves ($f$-mode), which
are well suited for studying the surface and sub-surface structure
of solar convection on supergranulation scales
\citep{2000SoPh..192..177D}. Our analysis shows that travel times
from the simulation agree very well with travel times from the
SOHO/MDI observations. This demonstrates that the simulated data are
sufficiently close to the observed solar data, providing ground for
further detailed helioseismic measurements and inversions.
Large-scale structures can be detected in the simulated data set
using $f$-mode time-distance and local correlation tracking
techniques and are observed directly in filtered flow fields.

\section{Numerical Model and Observed Data} \label{model}
We study the results of a 3D, compressible, radiative-hydrodynamic
(RHD) code simulating the upper solar convection zone and
photosphere. The code calculates LTE, non-gray radiation transfer
and employs a realistic equation of state and opacities. More
description of the code is found in
\citet{{2000SoPh..192...91S},{Benson2006}}.

The computational domain of our simulations spans 48 Mm $\times$ 48
Mm horizontally and 20 Mm vertically, with a horizontal resolution
of 100 km and vertical resolution of 12 to 75 km.  The three
velocity components measured at 200 km above $\tau_{\rm cont} = 1$
are saved every minute. Their spatial grid is 500 by 500 pixels. The
domain is sufficiently large, and the time sequence of several hours
is long enough to obtain sufficient signal/noise by temporal
averaging to perform a helioseismic time-distance analysis and seek
evidence for the presence of large-scale flows. Eventually, the
depth dependence and dynamics of these large-scale flows will be
studied as well. We compare some of our findings with results
obtained from SOHO/MDI high resolution observations. This
observational data set is an 8.5 hour time series, with 1 minute
cadence, of a 211.5 Mm $\times$ 211.5 Mm horizontal patch of the MDI
Doppler velocity, on the spatial grid of 512 by 512 pixels with
the resolution of about 400 km per pixel. The solar rotation is
removed. Both sets of data are processed identically, whenever a
comparison between the observations and simulations occurs.

\section{Surface Structures}
\label{structures}

Oscillation modes in the simulations are excited naturally, due to
convective motions and realistic cooling at the solar surface
\citep{2000ApJ...530L.139G,2001ApJ...546..585S,2004SoPh..220..229S}.
We use the vertical velocity component from the simulation data as a
proxy for the observed Doppler velocity, and calculate the
power spectrum, $P(k,\omega)$, in the horizontal
wavenumber-frequency domain. Following the helioseismology
convention, we represent the power spectrum in terms of the
spherical harmonic degree, $\ell=kR_\sun$, where $R_\sun$ is the
solar radius, and the cyclic frequency, $\nu=\omega/2\pi$.

\begin{figure}[ht]
    \plotone{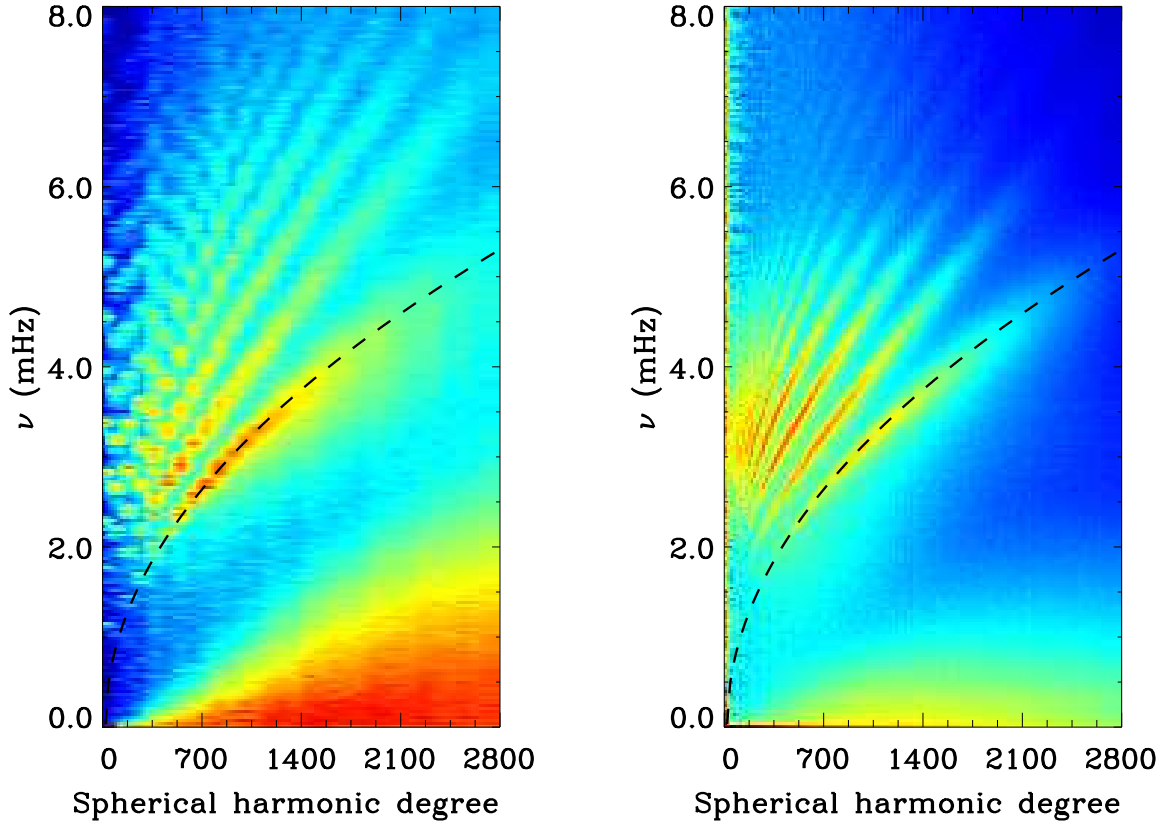}
    \caption{
    The power spectra ($\ell-\nu$ diagrams) for the simulated vertical velocity (left)
    and the Doppler velocity from the MDI high-resolution observations
    (right). The dark curve represents the theoretical $f$-mode.
    \label{KW}}
\end{figure}

The power spectrum ($\ell-\nu$ diagram) calculated from the
simulated vertical velocity clearly shows a resolved $f$-mode along
with a $p$-mode spectrum that looks very similar to the results from
the SOHO/MDI Doppler images (Fig.~\ref{KW}). These simulations
possess a richer mode spectrum than in our smaller (6 Mm wide by 3
Mm deep) simulations because the wider and deeper domain encompasses
many more resonant modes within it.  The spectral resolution
achieved in these simulations allows us to test time-distance
helioseismology measurements
\citep{{1993Natur.362..430D},{1997SoPh..170...63D}}.

\begin{figure}[!ht]
        \plotone{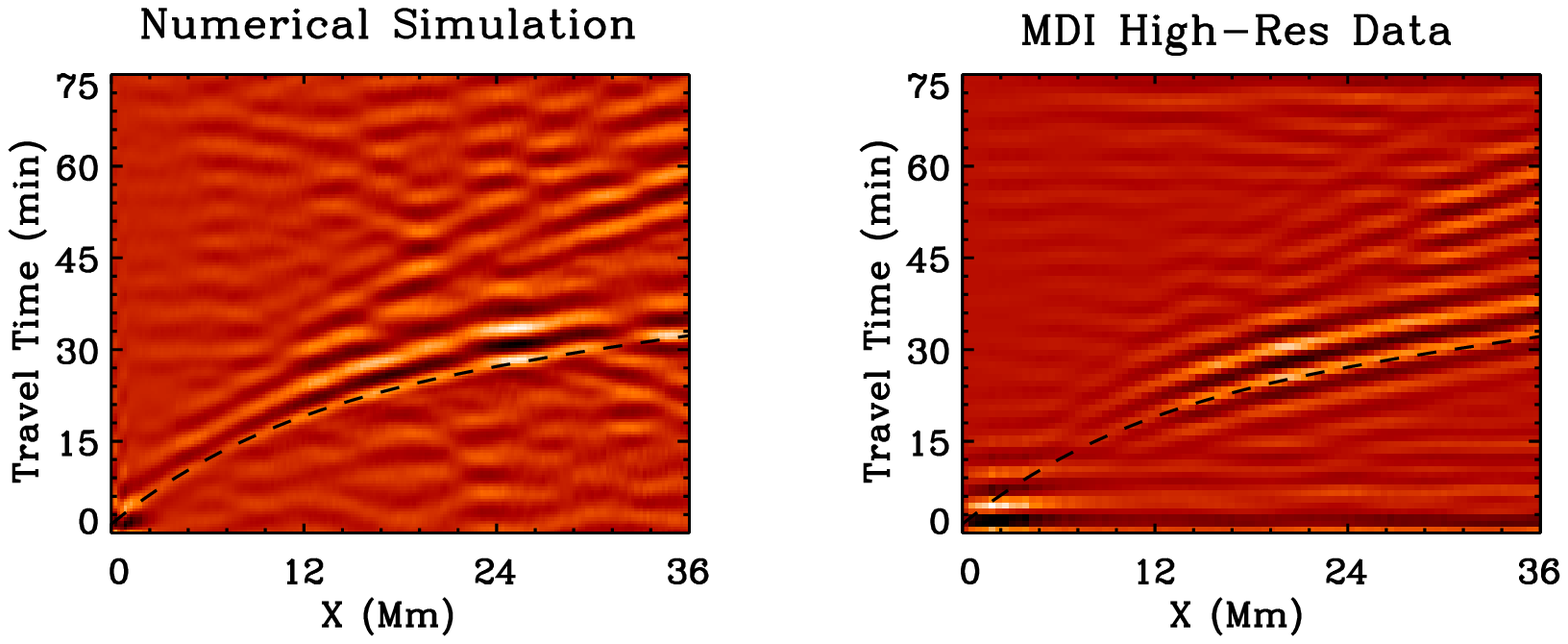}
        \caption{
    Time-distance diagrams for the simulated vertical velocity (left)
    and the Doppler velocity from the high-resolution MDI observations
    (right). The dark curve is time-distance relation computed from a
    standard solar model. The second bounce can be seen in both pictures.
        \label{TD}}
\end{figure}

Mode filtering is an essential part of the time-distance technique.
The convection signals (broad wedge in the lower right part of the
$\ell-\nu$ diagram) need to be removed. 
We construct the time-distance diagram for both
simulated and observed data by computing cross-covariance of
oscillation signals separated by a certain distance (Fig.~\ref{TD})
\citep{1997SoPh..170...63D}, and find a remarkable similarity
between them. Because of the higher resolution, in the
numerical simulation the cross-covariance signal extends to much
shorter distances than in the MDI data. This illustrates a
potential for small-scale helioseismic diagnostics of observations
of higher than MDI resolution, such as anticipated from the Solar-B
mission \citep{2004ASPC..325...87S}.

We show in section \ref{fmode} that the travel times from simulation
data agree with travel times from the SOHO/MDI observations very
well. Both the $\ell-\nu$ and time-distance diagrams illustrate the
potential of the simulations for testing helioseismology measurement
and inversions procedures.

In the following sections, we explore the surface structures in
the simulated data,  directly by averaging the flow field and using
spectral analysis (Sect.~\ref{direct}), and also by implementing the $f$-mode
time-distance analysis (Sect.~\ref{fmode}), and local correlation
tracking technique (Sect.~\ref{lct}).

\subsection{Fourier Analysis of Surface Velocities}
\label{direct}

We analyze the simulated velocity fields by calculating
spatial power spectra for both vertical and horizontal velocity
components and compare these with the spatial
power spectrum for the MDI Doppler signal.
\begin{figure}[!t]
        \plotone{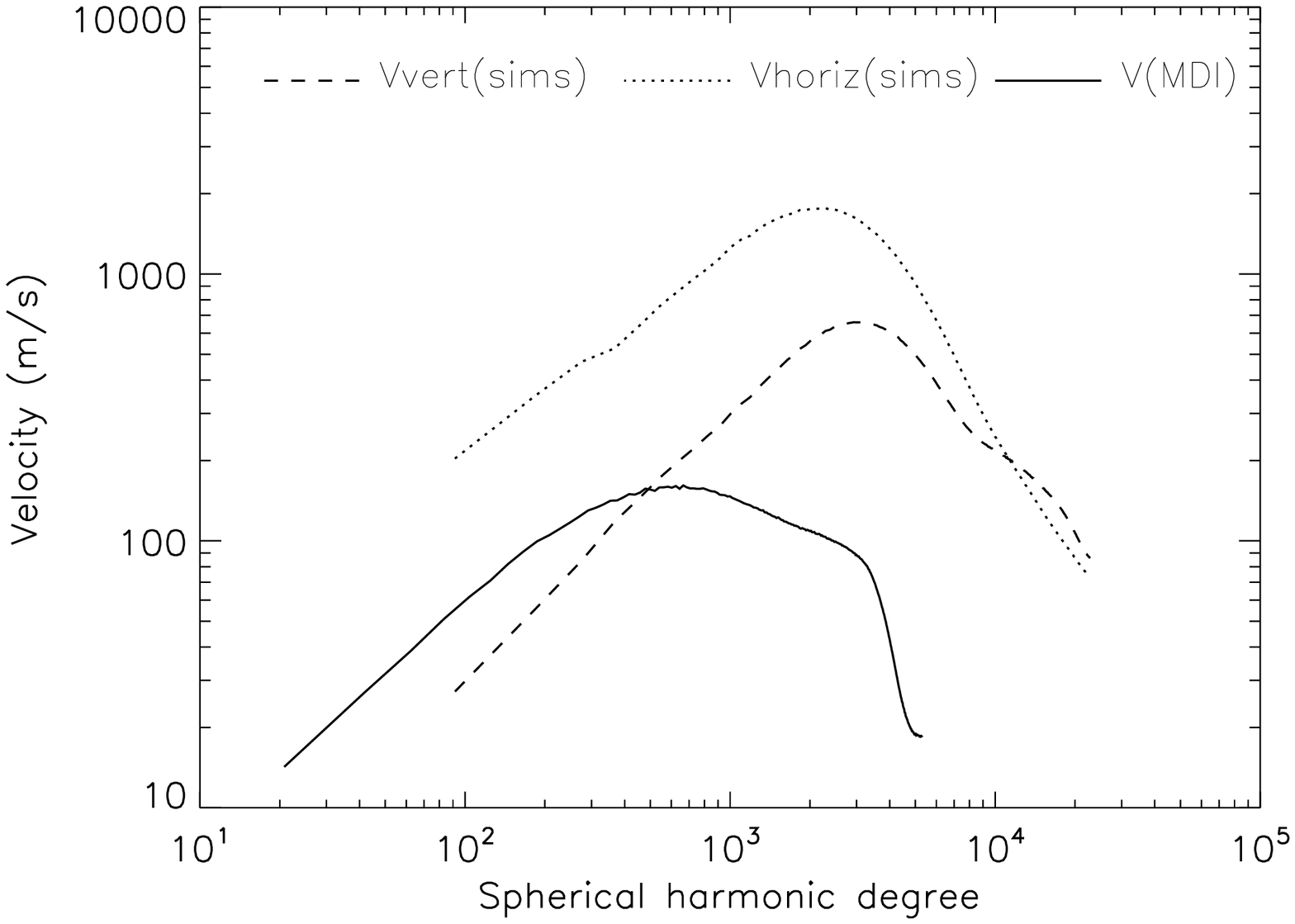}
        \caption{
    Spatial spectra for horizontal (dotted) and vertical (dashed) components of the
    simulated velocity, and the MDI Doppler signal (solid). Spectra are
    calculated for individual snapshots and then time-averaged, the
    simulation data over 20 hours and the MDI data over 8.5 hours.
        \label{SS}}
\end{figure}
Power spectra are calculated for each time and are then averaged
over time.  This reduces the statistical fluctuations that 
are present in spectra from a single time.
In Fig.~\ref{SS} we show $V(\ell) = (\ell \, P(\ell))^{1/2}$, where 
$V(\ell)$ is velocity amplitude in the Fourier domain at spherical 
harmonic degree $\ell$ and $P(\ell)$ is time averaged power (velocity
squared) per unit $\ell$. There is a continuous decrease in the
velocity amplitude from granules ($\sim$ 1.5 Mm, $\ell \sim$ 2000-3000) 
to larger scales. Aside from
granulation no special scales are revealed in the spectra, although
there may be a small enhancement in the horizontal velocity ($V_{\rm
horiz}$) spectrum at supergranulation scales ($\ell \sim 200$).

\begin{figure}[!t]
        \plotone{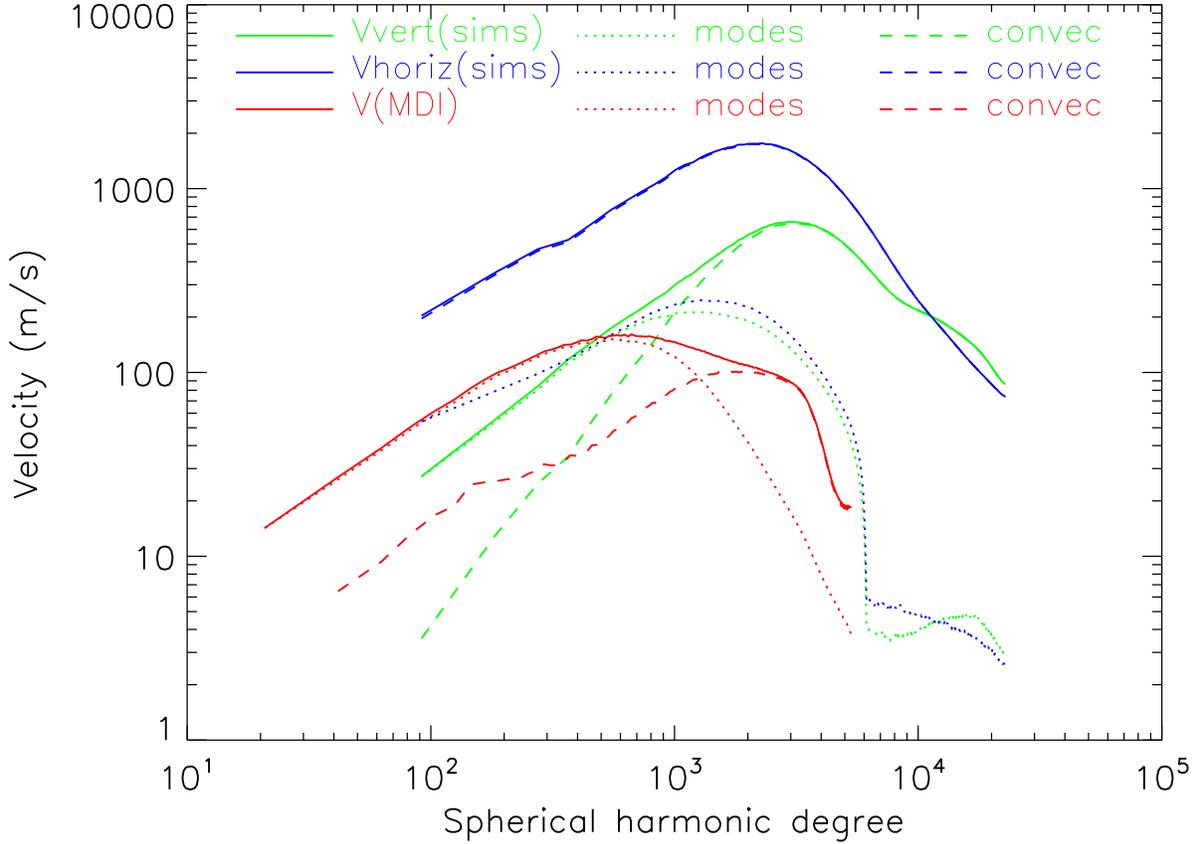}
        \caption{Velocity spectra for the simulation (vertical = green, 
	  horizontal = blue) and MDI observations (red) separated into 
	  convective (dashed lines), oscillatory (dotted lines)
      and total (solid lines) components. The oscillatory signal is negligible in
      the horizontal velocity but dominates the vertical velocity and
      MDI Doppler velocity on large scales ($\ell \le 1000$).
      \label{V_conv_osc}}
\end{figure}

In Fig.\ \ref{V_conv_osc} we separate motions into `convective' (dashed
lines)
and `oscillatory' (dotted lines) parts, according to whether the power in Fourier 
space lies above or below the line $\omega = c k$, where 
$c=6$ km\,s$^{-1}$ is approximately equal to the sound speed 
at the photosphere.  As can be seen from Fig.~\ref{KW} this provides
a rather clean split; the corresponding line in Fig.~\ref{KW}
would intersect the right hand side $\nu$-axis at $\nu\approx3.8$ mHz.

As shown by Fig.\ \ref{V_conv_osc}, the horizontal velocities (blue
curves) at the height of the MDI Ni $\lambda$ 676.78 nm line formation
(about 200 km above $\tau_{\rm 500} = 1$) are almost exclusively
of convective origin.  The vertical velocities (green curves), on the other hand,
are a mixture of convective (dashed line) and oscillation (dotted line) motions where the
convective motions are dominant at smaller scales and the oscillatory
motions are dominant at larger scales.

For granular and larger scales the
horizontal velocities are larger than the vertical velocities and
become increasingly dominant as the scale increases.  The reason that 
large scale horizontal motions are present at these heights is that
the atmosphere is not far from hydrostatic equilibrium and the pressure
fluctuations that exist below the surface to drive the large scale
horizontal flows imprint their fluctuations on the surface and produce
smaller amplitude large scale horizontal flows there also 
\citep{1982A&A...107....1N}.

Because of mass conservation, vertical convective velocities decrease 
more rapidly with size than the horizontal ones, with the ratio of
vertical to horizontal velocity amplitudes approximately inversely
proportional to size (cf.\ the green and blue dashed lines in 
Fig.\ \ref{V_conv_osc}).  At granular scales, where the vertical
velocity peaks, the vertical convective velocity at the height of 
formation of the $\lambda$ 676.78 nm line is still about 3--4 times
weaker than the horizontal one.

\begin{figure}[!ht]
         \plotone{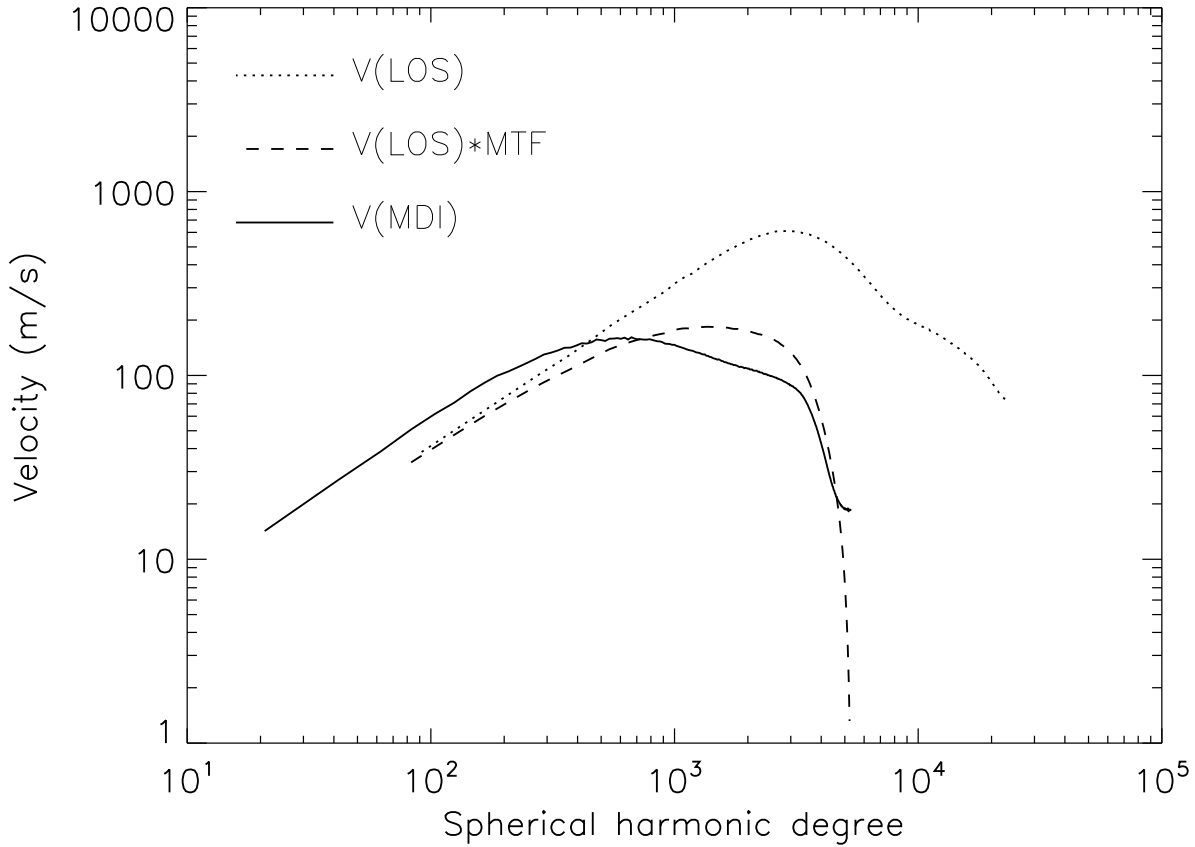}
         \caption{Simulated line-of-sight (LOS) component
         of the surface velocity spectrum (dotted), convolved with
      the instrumental MTF (dashed), compared with the MDI
         high-resolution spectrum (solid).
       \label{kw_MTF}}
 \end{figure}

In general, the MDI Doppler signal is a mix of horizontal and
vertical velocities. There is no obvious way to decouple these
components, although, depending on where on the solar disc the data
patch was located, one can estimate the approximate contribution of
the horizontal and vertical components. The current dataset was taken
near solar disk center (which at the time was about 7.5 heliographic 
degrees below the solar equator).  Nevertheless, because of the 
significant extension of the patch, the root-mean-square projection
of horizontal velocities into the line-of-sight was about 15\%,
while almost 99\% of the vertical velocity is projected onto
the line-of-sight. Indeed, if we combine 15\% of the
simulated horizontal velocities and 99\% of the simulated vertical
velocities, multiply by the instrument Modulation Transfer Function
(MTF) \citep{2001ApJ...548L.103W} and compare the resulting power
spectrum with the MDI power spectrum in Fig.~\ref{kw_MTF}, we get a
good correspondence, although the simulated power at large
scales (mostly oscillatory according to Fig.\ \ref{V_conv_osc}) falls 
somewhat short of the observed power.  Given that the numerical resolution
(96 km) is somewhat marginal on granular scales, where solar oscillations
are excited \citep{2001ApJ...546..585S}, this is not surprising.

From the slope of the MDI convective velocity (red dashed) curve in Fig.~\ref{V_conv_osc} one 
surmises that the convective LOS velocity observed with MDI over this 
patch actually derives most of its contribution at large scales from 
the horizontal velocity field.  An enhancement at supergranular scales 
(15-55 Mm, $\ell \sim$ 80-300) is visible.

\begin{figure}[!ht]
        \plotone{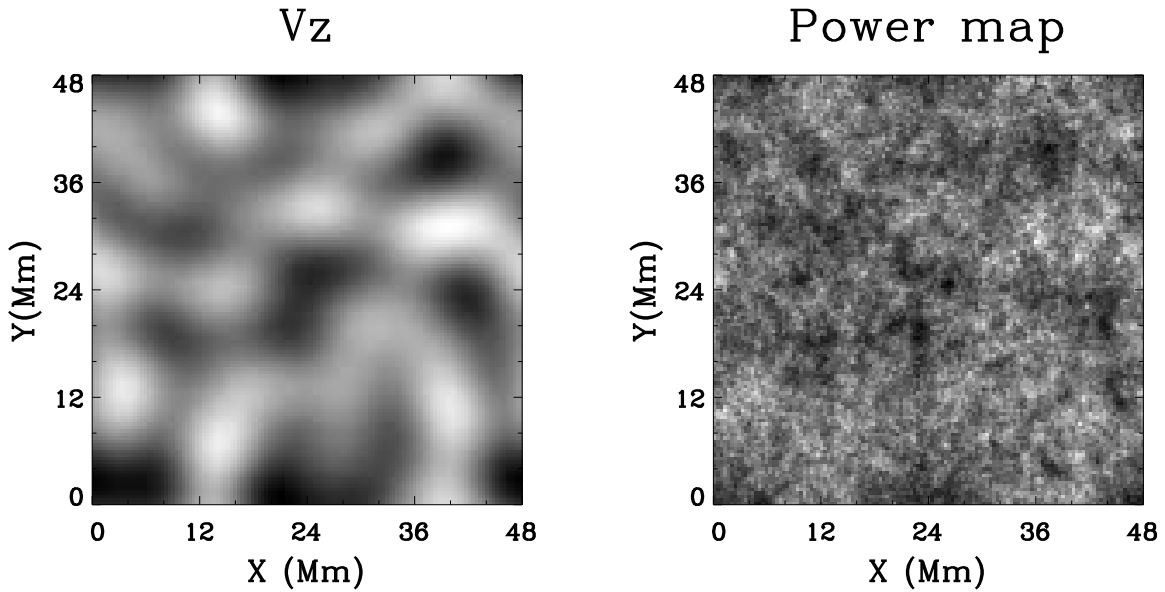}
        \caption{
        Left: Simulated large scale vertical velocity. The velocity is
      averaged over 8.5 hours and smoothed with a low-pass filter.
      Light is downflow and dark upflow.
      Right: $p$-mode power in the frequency interval 2-5 mHz.  No
      low-pass filter has been applied.  The oscillatory power is
      concentrated in the intergranular lanes and is weakest in the
      upflow centers of the large-scale structures.
        \label{VzPmap}}
\end{figure}

Averaging the velocity fields over several hours
eliminates the oscillations and suppresses the small scale features
but does not eliminate them.  To remove the small spatial scales, it
is necessary, in addition, to apply a low-pass k-space filter, 
where all values outside the circle $\sqrt{(k_x^2+k_y^2)} < 0.1$ Mm$^{-1}$ 
are set to zero.
We use this procedure to reveal the large-scale patterns in both
vertical and horizontal velocity components. Fig.~\ref{VzPmap} (left
panel) is an example of such a pattern in the $V_z$ component. An
alternative, cleaner, procedure would be to filter entirely in Fourier space,
eliminating the oscillations by removing power for $\omega \geq c k$ 
(as discussed in connection with Fig.\ \ref{V_conv_osc} above) and 
then applying a low-pass filter in $k$.  Here we follow the first
procedure which has been generally used in time-distance helioseismology.

We also construct $p$-mode power maps by Fourier-transforming the
vertical velocity fields in time at each spatial point, and summing
the result over the frequency interval 2-5 mHz (Fig.~\ref{VzPmap},
right panel). The $p$-mode power is rather intermittent on small
scales, and is in fact concentrated in intergranular lanes, as
shown by \citet{2001ApJ...546..585S}.  A large scale modulation,
correlated with the large scale vertical velocity field, is also 
discernible.  Presumably the large scale modulation is caused by
the slightly larger velocity amplitudes present in intergranular
lanes inside supergranulation scale downdrafts.

\subsection{F-Mode Time-Distance Analysis}
\label{fmode}

\begin{figure}[!ht]
        \plotone{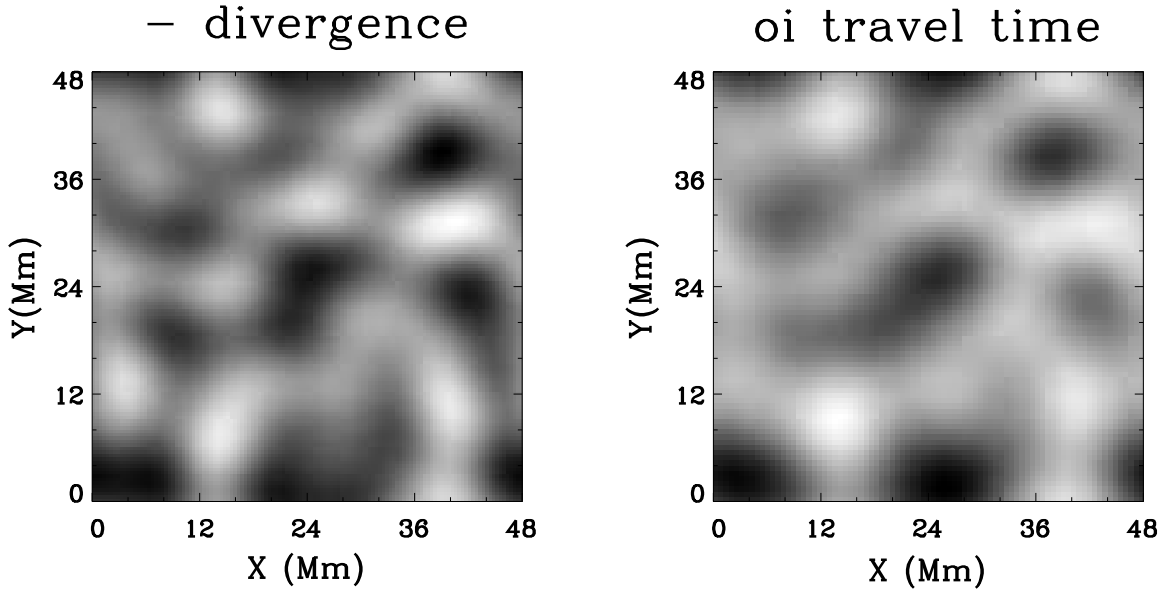}
        \caption{
    Comparison of the simulated horizontal velocity divergence (left;
    converging flows are light, diverging flows are dark) with
    $f$-mode outgoing and ingoing travel time difference map (right;
    incoming waves are faster in light areas, while outgoing waves are
    faster in dark areas)
    after a low-pass filter (as in Fig.~\ref{VzPmap}) has been applied.  Travel time
    extrema are 50 s for outgoing and 73 sec for incoming waves.
\label{OI}}
\end{figure}

\begin{figure}[!ht]
        \plotone{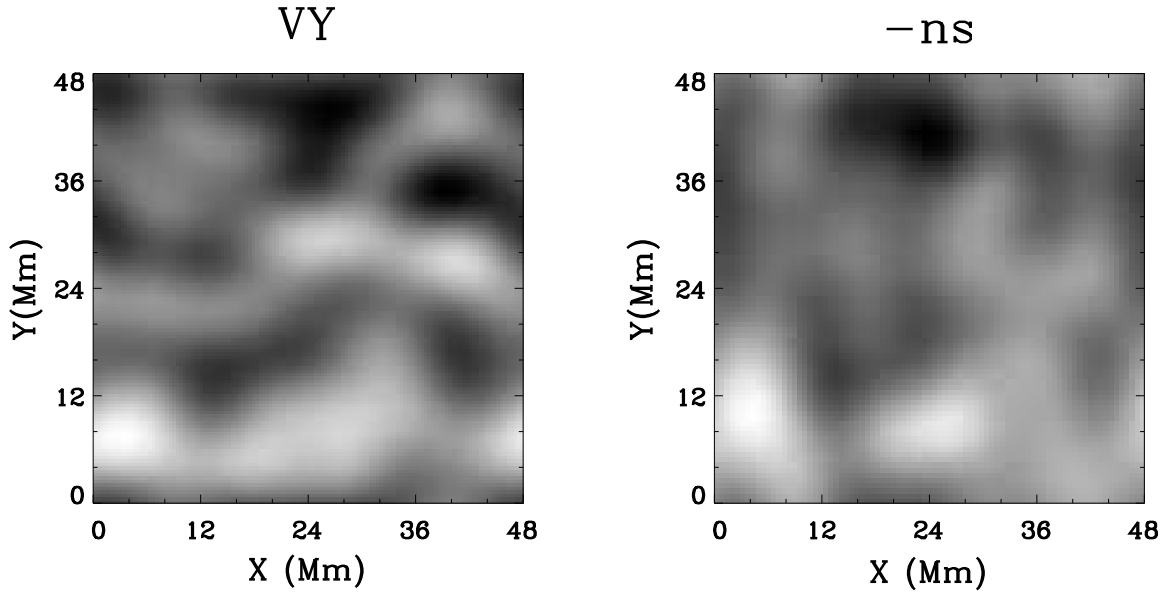}
        \caption{
        Comparison between the simulated horizontal velocity ($V_y$,
	  left, light is north-bound and dark is south-bound) and
        $f$-mode north-south travel time difference map (right, light
	  represents faster north-bound and dark represents faster 
	  south-bound waves). A low-pass
        filtering has been performed here as well. Travel time extrema
	  are 39 s for north-bound and 72 s for south-bound waves.
\label{NS}}
\end{figure}

An important question is whether local helioseismology methods can recover the
flow pattern from the oscillatory component of the
surface velocity field in these simulations.
We applied the time-distance technique to the $f$-mode oscillation,
because $f$-mode is the surface gravity mode and thus contains information
about horizontal flows and structures near the surface.
Therefore, the time-distance results can be directly compared with the
surface properties. The $f$-mode analysis is commonly used in
helioseismology.
We isolate the $f$-mode in $k_x, \, k_y, \, \nu$ space by retaining 
only values along a band centered on the theoretical $f$-mode ridge 
and setting all other values to zero, and then perform the time-distance
analysis, constructing outgoing and ingoing travel time maps, mean travel
times and travel time differences. We calculate the horizontal divergence
$dV_x/dx + dV_y/dy$ from the simulation data and compare it to the travel
time differences map after a low-pass filtering (Fig.~\ref{OI}).
As shown by \citet{2000SoPh..192..177D}, the travel-time difference is 
roughly proportional to the horizontal flow divergence.
We see a good agreement in the large-scale structures, with
correlation coefficient 0.88. Calculating
east-west and north-south travel time differences, we obtain proxies
for horizontal velocity components $V_x$ and $V_y$. We compare these
time difference maps with the actual velocities (Fig.~\ref{NS} for the
North-South componenet, the other looks similar). After
low-pass filtering, there is a good qualitative agreement between the two,
with correlation coefficients 0.7 for $x$-component and 0.73 for $y$-component.
It is worth mentioning that the time-distance technique measures an
average over a certain depth, therefore a very high correlation is
not to be expected when comparing the time-distance results to a
single depth data, particularly above the surface. The agreement
might have been better if one looked at a depth representative of
the $f$-mode travel paths ($f$-mode kinetic energy is concentrated 
in the 2 Mm just below the surface).

\subsection{Local Correlation Tracking}
\label{lct}

\begin{figure}[!ht]
    \plotone{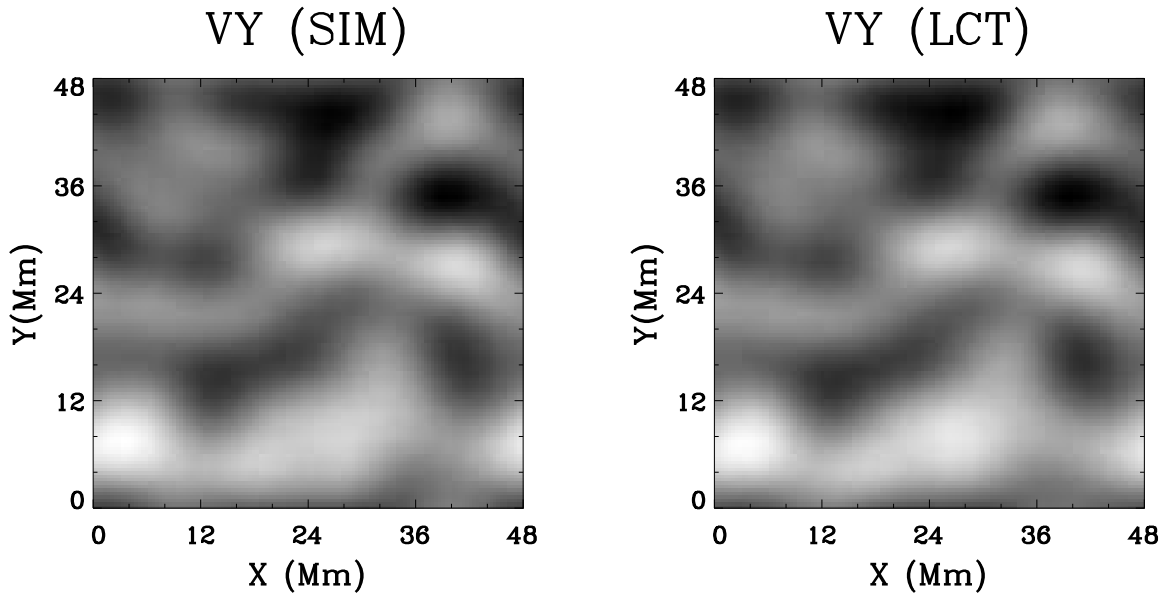}
    \caption{
    Simulated horizontal velocity component $V_y$ (left), averaged
        over 8.5 hours, and $V_y$, obtained by LCT analysis (right), after
        low-pass filtering.
        \label{Vy_LCT}}
\end{figure}

Another possibility for deducing the horizontal velocities is to apply
the local correlation tracking (LCT) technique to the vertical velocity.
The LCT is a cross-correlation method
\citep[e.g.][]{1988ApJ...333..427N} applied to a time series
of solar granulation images. The cross correlation is spatially localized (within a
certain window, usually a Gaussian); its time average presumably measures
horizontal displacements of the flows.
We apply the LCT technique to the simulated vertical velocity field
and compare the resulting horizontal velocity proxies with the simulated
horizontal velocity components.  
Fig.~\ref{Vy_LCT} shows the $V_y$ component.
Velocities determined by the LCT method are tightly correlated with the
simulation velocity, 
with correlation coefficients of 0.99 for both $V_x$ and $V_y$ components.  
However, the LCT amplitudes are a factor 1.8 smaller.
Application of a low-pass filter to the simulation velocities brings this 
ratio down to 1.5.  These results are consistent with conclusions of 
\citet{2001A&A...377L..14R}.  According to them, the horizontal velocity
proxies deduced from granular motions underestimate actual velocities
by a variable factor, from 2.1 at small scales to 1.6 at large scales.

Thus, the realistic simulation data also provide a tool for testing
and improving the local correlation tracking techniques.

\section{Discussion}
\label{conc}

The new large-scale realistic 3D radiative-hydrodynamic simulations
of the upper layers of solar convection provide an excellent
opportunity to validate various techniques, widely used in solar
physics and helioseismology for directly obtaining otherwise
inaccessible properties (subsurface flows and structures, etc). On the
other hand, these analysis techniques also help to examine how
realistic the simulations are. We have performed an initial analysis
of the simulated data and compared our results with the outcome of
the SOHO/MDI observations. The similarity between the simulated and
observed $\ell-\nu$ and time-distance diagrams demonstrates that the
simulations can be efficiently used to perform and validate local
helioseismology techniques. 
This agreement also reveals a potential for high-resolution
time-distance measurements.
We carried out $f$-mode time-distance
calculations and local correlation tracking to obtain horizontal
velocities. We compare them to the simulated horizontal flows and
find a good qualitative agreement. We see large-scale structures in
both actual horizontal velocities and their proxies. The results of
this investigation provide the basis for further detailed
helioseismic analyses of the simulated data, including various
scheme of measuring p-mode travel times, approximations for the
sensitivity kernels and inversion procedures.

{\bf Acknowledgements.}  DG is grateful for the support of the SOHO/MDI
group at Stanford University.  DB and RFS were supported by NASA
grants NNG04GB92G and NAG 512450 and NSF grants AST-0205500 and
AST-0605738.  The work of {\AA}N was supported by a grant from the
Danish Natural Science Research Council.  Computing time was provided
by the Danish Center for Scientific Computing, and by grants from
the NASA Advanced Supercomputing Division.

\bibliography{ms}

\end{document}